\documentclass[sigconf]{acmart}
\usepackage{booktabs,graphicx,tikz,pgfplots}
\usetikzlibrary{arrows.meta,positioning,calc}
\pgfplotsset{compat=1.18}
\definecolor{blue}{HTML}{246486}
\definecolor{teal}{HTML}{348477}
\definecolor{orange}{HTML}{B96A35}
\newcommand{\score}{\operatorname{score}}
\newcommand{\vlen}{\operatorname{len}}

\usepackage{balance}

\copyrightyear{2026}
\acmYear{2026}
\setcopyright{cc}
\setcctype{by}
\acmConference[SIGSPATIAL '26]{The 34th ACM International Conference on Advances in Geographic Information Systems}{November 03--06, 2026}{Riverside, CA, USA}
\acmBooktitle{The 34th ACM International Conference on Advances in Geographic Information Systems (SIGSPATIAL '26), November 03--06, 2026, Riverside, CA, USA}
\acmDOI{10.1145/3841645.3850172}
\acmISBN{979-8-4007-2950-8/2026/11}

\title{Simulated Annealing for Antenna Placements}
\author{Jacobus Conradi}
\affiliation{\institution{Department of Computer Science, University of Copenhagen}\city{Copenhagen}\country{Denmark}}
\email{jaco@di.ku.dk}
\author{Asbj{\o}rn Lind}
\affiliation{\institution{Department of Computer Science, University of Copenhagen}\city{Copenhagen}\country{Denmark}}
\email{asbjorn.lind@di.ku.dk}
\author{Mikkel Abrahamsen}
\affiliation{\institution{Department of Computer Science, University of Copenhagen}\city{Copenhagen}\country{Denmark}}
\email{miab@di.ku.dk}

\begin{document}
\begin{abstract}The 2026 ACM SIGSPATIAL GIS Cup posed the following geometric challenge: Given a set of simple, pair-wise disjoint polygons, compute \(k\) antennas (points) on polygon boundaries, maximizing the number of polygons with at least a fraction \(\tau\) of their perimeter visible from the antennas. Contestants had 24 hours to produce the best possible solutions.

We describe an approach that separates geometric visibility preprocessing from a portfolio of incremental combinatorial searches. A construction pool provides initial solutions with antennas restricted to polygon vertices; simulated annealing explores one-for-one antenna swaps, complemented by exact discrete one-swap descent, ruin-and-recreate, and elite crossover. A secondary objective rewards progress toward the service threshold without overriding the primary score. Candidate antennas are later expanded to include points in polygon-edge interiors. Our submitted solutions cover between 261 and 16,273 buildings across the nine parameter combinations, and the team was invited to present as one of the top entries. The code is available at \url{https://github.com/JacobusTheSecond/giscup}.\end{abstract}
\ccsdesc[500]{Information systems~Geographic information systems}
\ccsdesc[300]{Theory of computation~Computational geometry}
\keywords{Antenna Placement, Visibility, Simulated Annealing, Local Search, GIS Cup}
\maketitle
\section{Introduction}

The GIS Cup is an annual computational challenge hosted by ACM SIGSPATIAL~\cite{giscup}. The 2026 problem combines geometric visibility with a discrete service objective. Given a set of polygons $B$, referred to as \emph{buildings}, we must place exactly $k$ points, or \emph{antennas}, on building boundaries. A boundary point is \emph{visible} from an antenna if the connecting segment does not cross the interior of any building. A building is \emph{serviced} when the union of its visible boundary portions covers at least a fraction $\tau$ of its perimeter. The goal is to maximize the number of serviced buildings for given $k$ and $\tau$.

This objective creates a difficult search landscape. Additional visible perimeter has no value until a building crosses the threshold $\tau$, while a small loss can remove it from the score. Antennas can also complement one another: two antennas may jointly service a building even when neither does so individually. For the given instance, this is often necessary, especially when $\tau\geq 0.5$. Greedy additions and strictly improving single swaps can therefore become trapped on broad plateaus. In related computational-geometry challenges, members of our team used local improvement for non-obtuse triangulations~\cite{nonobtuse} and simulated annealing for median triangulations~\cite{median}. We use related techniques here.

The competition dataset contained $50\,000$ buildings and nine parameter combinations,
$\tau\in\{0.32,0.49,0.68\}$ and $k\in\{9,49,484\}$, released 24 hours before the submission deadline~\cite{giscup}.

\section{Problem Representation and Search Framework}

\subsection{Coverage objective}

For an antenna $a$ and building $b\in B$, let $V_a(b)\subseteq\partial b$ denote the visible portion of its boundary. For a set of antennas $P$,
\begin{equation}
C_P(b)=
\frac{\vlen\bigl(\bigcup_{a\in P}V_a(b)\bigr)}
     {\vlen(\partial b)},
\qquad
\score(P)=
\sum_{b\in B}\mathbf{1}[C_P(b)\geq\tau].
\end{equation}
The constraint is $|P|=k$.

The score does not distinguish between different amounts of partial coverage. A building just below its threshold contributes exactly as much to the score as one with almost no visible boundary. We therefore use additional guidance during the search while keeping the number of serviced buildings as the primary objective.

\begin{figure}[t]
\centering
\includegraphics[width=0.45\textwidth]{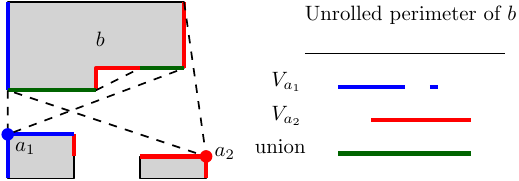}
\caption{Visibility and perimeter coverage (schematic). Antennas lie on boundaries; dashed lines highlight visibility.}
\label{fig:coverage}
\end{figure}

\subsection{Overall approach}

We separate the expensive geometric computations from the combinatorial search. Visibility is first computed for the discrete set of candidate antenna positions, defined by the vertices of all buildings, and stored in a compact form. We then construct several solutions with $k$ antennas and improve them using simulated annealing, deterministic one-swap search, ruin-and-recreate, and crossover. In a later phase, we add candidate positions in the interiors of building edges and continue the search.

Several searches run independently and may temporarily accept worse solutions, while the best solution found so far is retained separately.

\subsection{Cached visibility}

The initial candidate set consists of building vertices. We use CGAL geometric primitives to compute visibility~\cite{cgal}, intersect the resulting visible regions with building boundaries, and encode the resulting boundary intervals together with their lengths. After this preprocessing, evaluating a search move only requires these intervals and the association between candidates and the intervals they see; no new visibility computation is needed.

When one antenna is replaced by another, only buildings seen by either the removed or inserted antenna can change their coverage. We therefore update only those buildings. Each independent search maintains its own coverage state, and cached local information is discarded when a search is restarted from another solution.

\section{Construction and Fixed-Cardinality Search}

\subsection{Construction with an effective threshold}
\label{sec:construction}

We first build a solution by adding antennas one at a time until $k$ have been selected. Using the final threshold $\tau$ from the beginning gives little guidance: with only a few antennas, almost no building may be close to satisfying the final requirement.

Instead, at cardinality $j\leq k$ we use the \emph{effective threshold}
\begin{equation}
\tau_j=
\tau\left(
\frac{\lceil sj/k\rceil}{s}
\right)^m,
\qquad
s=100,\quad m=0.75.
\end{equation}
As more antennas are added, $\tau_j$ approaches the final threshold and satisfies $\tau_j=\tau$ at $j=k$. We chose this particular form of the effective threshold for two reasons. Consider the simpler form $\tau(j/k)$. At every increase $j\gets j+1$, this threshold changes, which triggers a set of recomputations and invalidations in the cache. With the 24-hour time limit, this is undesirable. Stepping the threshold instead, via $\tau\lceil sj/k\rceil/s$, bounds these invalidations to at most $s$, while approximately maintaining the same behavior. Raising $\lceil sj/k\rceil/s$ to the power $m$, with $m<1$, makes us spend more of the construction time at thresholds closer to the final threshold $\tau$.

Several solutions are constructed in parallel rather than following a single greedy sequence. At regular time intervals, partial solutions are combined, and greedy one-swaps are considered. We allocate roughly $6$ hours for the entire initial construction. The completed solutions become starting points for the fixed-cardinality search.

\subsection{Simulated annealing}
\label{sec:annealing}

The overarching idea of simulated annealing is as follows: one maintains a parameter $T$, typically called the \emph{temperature}. The higher the temperature, the more explorative the algorithm is, and the lower the temperature, the more exploitative. At any time, one has a solution to the problem and considers a set of candidate modifications. One modification is then selected depending on its quality and the current temperature.

In our case, the basic move modifying a solution is a one-swap: remove one selected antenna and insert one unselected candidate antenna. At each iteration, we consider a batch of insertion candidates and choose a promising removal for each. A batch contains 1,024 insertion candidates.

\subsection{Evaluation}
\label{sec:guidance}

Given a set of antennas, its score can be too coarse to guide the search effectively. Consider, for example, two solutions with the same score, where one has many buildings at $0.99\tau$ service and the other has no such buildings. Naturally, the first solution should be considered more promising. To capture this effect, we introduce the \emph{quality} of a solution.

For this, first define how close a building $b$ is to being serviced:
\(
r_b=\min\{C_P(b)/\tau,1\}
\).
With $f=0.5$, we use
\begin{equation}
h(r)=
\begin{cases}
0.05r/f, & 0\leq r\leq f,\\
0.05+0.95\bigl((r-f)/(1-f)\bigr)^2, & f<r\leq1,
\end{cases}
\end{equation}
and define
\(
H(P)=|B|^{-1}\sum_{b\in B} h(r_b)
\).
The function gives only a small reward to buildings far below the threshold and places most of the additional weight on buildings that are close to being serviced.

We also use the total fraction of building boundary that is visible,
\[
L(P)=
\frac{
\sum_{b\in B}
\vlen\left(\bigcup_{a\in P}V_a(b)\right)
}{
\sum_{b\in B}\vlen(\partial b)
}.
\]
Finally, the quality of a solution is
\begin{equation}
Q(P)=\score(P)+0.70H(P)+0.299L(P).
\end{equation}

Both $H(P)$ and $L(P)$ lie in $[0,1]$, so together they contribute at most $0.999$. An additional serviced building therefore always outweighs any change in the secondary terms.

The quality is the annealing objective. At any stage, a temporary solution considers a sampled set of one-swaps. These swaps are evaluated using the quality $Q$. With $\Delta Q=Q(P')-Q(P)$ the difference in quality of the unmodified to the modified solution, a selected move is accepted according to the standard simulated-annealing rule~\cite{kirkpatrick}, $\min\{1,\exp(\Delta Q/T)\}$. The temperature is calibrated from observed losses and then cooled geometrically. We run several independently seeded annealing searches and occasionally reheat them to increase the chance of escaping a local optimum.

Always pairing an insertion with its best removal can itself make the search too restrictive. Some higher-temperature searches therefore test several removals for each insertion, combining a few promising removals with randomly sampled ones.

\section{Additional Improvement Methods}

\subsection{Deterministic one-swap search}

In addition to simulated annealing, we periodically search exhaustively for an improving one-swap. For an insertion candidate $c$, the score of $P\cup\{c\}$ before removing an antenna is an upper bound on the score of every replacement
$(P\setminus\{r\})\cup\{c\}$. We use this bound to discard insertions that cannot lead to an improvement before evaluating individual removals.

If a complete scan finds no improving swap, the current solution is locally optimal with respect to the scanned candidate set. After an accepted move, only candidates related to buildings whose coverage changed need to have their cached bounds recomputed, which makes repeated scans considerably cheaper.

The deterministic search and simulated annealing serve different purposes. The deterministic scan efficiently finds remaining local improvements, while annealing can temporarily move to worse solutions. While simulated annealing is able to escape local maxima, which exhaustive searches cannot, neither directly handles improvements that require several coordinated replacements, which motivates the larger moves described next.

\subsection{Ruin-and-recreate and crossover}

Ruin-and-recreate removes several antennas from a solution and then fills the empty positions again, using techniques discussed in Section~\ref{sec:construction}. We use several removal strategies, including antennas with a small apparent contribution, spatially close antennas, antennas associated with buildings close to their threshold, random removals, and groups of antennas that see many of the same buildings.

Replacement candidates are evaluated against the partially destroyed solution. Removed antennas are temporarily excluded during most of the reconstruction to reduce the chance of immediately rebuilding the same solution. A reconstructed solution that does not improve the current best can still be used as a starting point for later annealing.

We also combine pairs of good solutions. Antennas occurring in both are retained, after which the remaining positions are filled again. This allows useful combinations of antennas to pass between independent searches without having to be rediscovered entirely through one-swaps.

\subsection{Edge refinement}

Vertices of buildings form an initial candidate set of manageable size. In our instance, the vertex-only search contains $306\,833$ candidate positions. The problem, however, permits antennas anywhere on building boundaries. A separate preprocessing job therefore generates additional candidate positions in the interiors of edges and computes their visibility. After removing near-duplicate positions, the expanded candidate set contains $11\,609\,644$ candidates, approximately $37.8$ times as many as the original vertex set.

Once these data are available, the search continues from the best solutions found during the vertex phase. Because the number of candidates is much larger, we do not evaluate every possible insertion. Instead, an insertion screen considers a small sample of the candidate set. At most 1,024 of the strongest candidates are then passed to the same selection procedure used during vertex search.

After edge search, selected non-vertex antennas can be adjusted along their source edge. We test movements of $1$\,mm in either direction and recompute visibility. A movement is accepted if it improves the service score or, at equal service score, $H(P)$. Vertex antennas are left unchanged.

Before this final adjustment, we reconstruct the current solution and verify that it services the same set of buildings. Similar checks are performed after score improvements and at the end of each phase. These checks help detect inconsistencies between the cached representation and the coordinate-level solution, but the submitted solution is ultimately evaluated by the organizers' implementation~\cite{evaluator}.

\section{Implementation and Competition Run}

The solver is implemented in C++ using CGAL and GDAL/OGR. Supplementary restricted-candidate experiments use OR-Tools. Competition jobs were run through Slurm on a cluster provided by the University of Copenhagen.

We ran one optimization job for each of the nine $(\tau,k)$ combinations. The three $k=484$ jobs used 96 CPU cores each, while the six $k\in\{9,49\}$ jobs used 64 cores each; every job requested 128\,GiB of memory. The nine jobs ran concurrently for 22\,h\,15\,min to 22\,h\,18\,min. At full concurrency this corresponds to 672 CPU cores and 1.125\,TiB of requested memory. Over their recorded runtimes, the worker jobs account for approximately 14,965 CPU-hours.

A shared preprocessing job generated the additional edge candidates and their visibility data using 48 CPU cores and 512\,GiB of memory. Including this job, the peak configured allocation was therefore 720 CPU cores and 1.625\,TiB of memory.

The main optimization was given 23 of the 24 competition hours. Vertex search was followed by edge search, with 45 minutes reserved for final coordinate adjustment. The submission archive records the code, configuration, submitted results, and selected solution for every parameter setting.

\section{Competition Results and Discussion}

\begin{figure*}[t]
\centering
\includegraphics[width=0.9\textwidth]{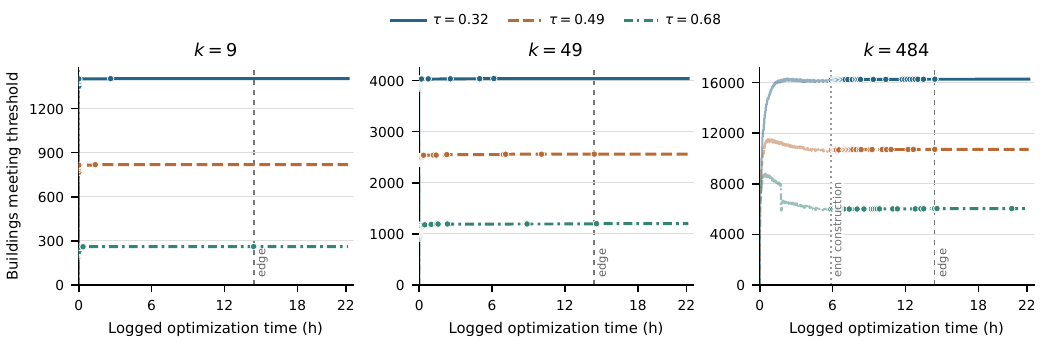}
\caption{Competition progress on a common timeline. Faint curves show construction under the effective threshold $\tau_j$; opaque step curves show the running best after reaching cardinality $k$, with circles marking improvements. Dotted and dashed lines mark the end of construction and start of edge search, respectively.}
\label{fig:progress}
\end{figure*}

Table~\ref{tab:scores} reports the score at the end of vertex search and the submitted score for each parameter setting.
\begin{table}[t]
\centering
\caption{Vertex-phase and submitted scores. $\Delta$ is the improvement during edge search.}
\label{tab:scores}
\begin{tabular}{rrrrr}
\toprule
$\tau$ & $k$ & Vertex & Submitted & $\Delta$\\
\midrule
0.32 & 9   & 1,403  & 1,403  & 0\\
0.32 & 49  & 4,039  & 4,039  & 0\\
0.32 & 484 & 16,269 & 16,273 & 4\\
0.49 & 9   & 818    & 818    & 0\\
0.49 & 49  & 2,559  & 2,561  & 2\\
0.49 & 484 & 10,716 & 10,724 & 8\\
0.68 & 9   & 260    & 261    & 1\\
0.68 & 49  & 1,194  & 1,199  & 5\\
0.68 & 484 & 6,025  & 6,043  & 18\\
\bottomrule
\end{tabular}
\end{table}

Figure~\ref{fig:progress} shows construction and later improvement on the same scale: the number of buildings meeting the threshold used at that point in the search. Construction was designed before the final values of $k$ were known, with a target duration of roughly six hours. In fact, our approach would have been able to handle up to $k=10\,000$. The released values of $k$ were considerably smaller than expected, however. Construction therefore finishes within seconds for $k=9$ and $k=49$, whereas for $k=484$ it takes 5.86--5.90\,h, close to the intended duration. At the end of construction, $\tau_j=\tau$, so the construction curve joins the final-threshold score directly.

All nine runs reach at least 99\% of their eventual submitted scores at the end of the construction phase. The remaining time is therefore spent finding relatively small improvements. For $k=484$, these improvements continue well after construction has ended.

The plotted time concatenates consecutive optimizer runs. The curves show the running maximum, so temporary decreases caused by exploratory moves are omitted. During construction, fewer than $k$ antennas may be present, so the faint curves show the construction score rather than a feasible final solution.

\paragraph{Edge search.}
Edge search improves six of the nine parameter settings, for a total increase of 38 serviced buildings. All three $k=484$ solutions improve, with the largest increase being 18 buildings for $(\tau,k)=(0.68,484)$.

The vertex solutions are already close to the submitted scores: each achieves at least 99.58\% of its final value. Nevertheless, some improvements occur late. For $(0.68,484)$, the final score is first reached after 20.75\,h.

These measurements show that improvements occurred during the edge phase, but they do not separate the benefit of adding edge candidates from the benefit of simply continuing the search. The score at the end of edge search also equals the submitted score in all nine cases, so we observe no further increase in the primary score during the final coordinate adjustment.

\paragraph{Additional experiments and limitations.}
For the six settings with $k\in\{9,49\}$, we also ran restart-heavy searches followed by CP-SAT optimization on restricted candidate sets. These match the main score for $(0.32,49)$ and $(0.49,9)$ and are worse by 2, 9, 3, and 5 buildings for the remaining four settings. None replaces a submitted solution.

These experiments neither isolate the contribution of the individual search methods nor establish optimality. The methods share starting solutions, and meaningful comparisons between stochastic methods would require controlled repeated runs.

A further limitation is the difference between the coverage representation used during search and the continuous geometry used by the official evaluator~\cite{evaluator}. The values reported here are the scores recorded for our submitted solutions and in the corresponding optimization logs.

\balance

\section{Conclusion}

We presented a solver that separates visibility computation from the combinatorial search for antenna positions. An effective threshold provides useful guidance during construction, while bounded secondary terms help simulated annealing move across score plateaus once $k$ antennas have been selected. Deterministic one-swap search, ruin-and-recreate, crossover, and a later expansion from vertices to edge candidates provide additional ways to improve a solution.

The main runs produced all nine submitted solutions. Most of the final score was reached early, but continued search still improved six parameter settings, including all three cases with $k=484$. Future work could generate edge candidates more selectively around buildings close to their service threshold and measure the contribution of the individual search methods using controlled repeated runs.

\end{document}